\documentclass[10pt]{article}
\usepackage{graphicx}
\begin{document}
\title{Duality symmetry conjugates of the quantum Rabi model : effective bosonic, fermionic and coupling-only dynamical properties}
\author{Joseph Akeyo Omolo\\
Department of Physics, Maseno University, P.O. Private Bag, Maseno, Kenya\\
e-mail:~ ojakeyo04@yahoo.co.uk}
\date{01 December 2021}
\maketitle

\begin{abstract}
Symmetry transformations have proved useful in determining the algebraic structure and internal dynamical properties of physical systems. In the quantum Rabi model, invariance under parity symmetry transformation has been used to obtain exact solutions of the eigenvalue equation and very good approximations of the internal dynamics of the interacting atom-light system. In this article, two symmetry operators, characterized as ``duality'' symmetry operators, have been introduced which transform the quantum Rabi Hamiltonian into duality conjugates. The parity and duality symmetry operators constitute an algebraically closed set of symmetry transformation operators of the quantum Rabi model. The closed $SU(2)$ Lie algebra provides the standard eigenvalues and eigenstates of the parity symmetry operator. It is established that Jaynes-Cummings and anti-Jaynes-Cummings operators are duality symmetry conjugates. Symmetric or antisymmetric linear combinations of the Rabi Hamiltonian and a corresponding duality conjugate yield the familiar spin-dependent force driven bosonic , coupling-only or quantized light mode quadrature-driven fermionic Hamiltonian. It is established that the effective bosonic, fermionic and coupling-only Hamiltonians are exact, not approximate forms of the quantum Rabi Hamiltonian as they have generally been interpreted. The effective bosonic form generates the dynamics of the light mode driven by the atomic spin-dependent force, while the fermionic form generates the dynamics of the atomic spin driven by the quantized light mode quadrature-dependent force, thus providing a complete picture of the quantum Rabi dynamics. Dynamical evolution of the full quantum Rabi model is thus described by entangled coherent light Schroedinger cat and atomic spin states of the effective bosonic or coupling-only Hamiltonian, or, by generalized atomic spin coherent states of the effective fermionic Hamiltonian. These nonclassical states of the simpler effective forms of QRM Hamiltonian have revealed fundamental quantum features observed in experiments and have been useful in developing various quantum technology applications.
\end{abstract}

\section{Introduction}
The quantum Rabi model (QRM) of a two-level atom interacting with a quantized light mode is generated by Hamiltonian
$$H=H_0+H_I \ ; \quad\quad H_0=\hbar\omega\hat{a}^\dagger\hat{a}+\hbar\omega_0s_z \ ; \quad\quad H_I=\hbar g(\hat{a}+\hat{a}^\dagger)(s_++s_-) \eqno(1a)$$
where $H_0$ , $H_I$ are the free evolution and interaction components. Here, $\omega$ , $\hat{a}$ , $\hat{a}^\dagger$ are the quantized light mode angular frequency, annihilation and creation operators, while $\omega_0$ , $s_z$ , $s_+$ , $s_-$ , $\sigma_x=s_-+s_+$ are the atomic state transition angular frequency and operators. The light mode vacuum state energy $\frac{1}{2}\hbar\omega$ has been ignored, but can be reintroduced as desired.

Due to its purely quantum nature, QRM has been the center of focus of both theory and experiments in quantum optics seeking to understand fundamental quantum mechanical properties and their potential applications to quantum technology development. The QRM Hamiltonian $H$ in equation (1$a$) takes a standard form applicable to a broad spectrum of physical processes based on light-matter interactions such as atoms or electrically charged particles in magnetic fields, cavity and circuit quantum electrodynamics, quantum dots, trapped ions, polaritonic physics, superconducting quantum circuits, etc, which are well described in the excellent reviews [1 , 2].

While great breakthroughs have been made in providing exact analytical solutions of the QRM eigenvalue equations generated by $H$ [3 , 4] based on the parity symmetry property, determining an exact general solution of the corresponding time evolution equation remains a major challenge of theoretical quantum optics. The spectrum of eigenvalues and eigenstates obtained in the exact solutions in the Bargmann space in [ 3 , 4] and related subsequent work not cited here turns out too complicated, meaning that some approximations still have to be applied in determining the time evolving state vector [5 , 6]. Faced with observable discrepancies between the experimental results [1 , 2 , 7-11] and the basic Jaynes-Cummings (JC) model obtained in the rotating wave approximation (RWA) [12 , 13 , 14], great effort has been made developing more sophisticated effective approximations to the QRM Hamiltonian $H$ [15-19]. Even though the results of these effective approximations are closer to experimental observations over the desired coupling ranges from weak to ultra/deep strong coupling values, some subtle features such as excitation-dependent damping rates are not yet properly captured.

Where, then, is the main problem of theory ? The answer to this basic question lies in the full quantum operator form of the QRM Hamiltonian $H$ in equation (1$a$). Since the quantized light mode and atomic spin operators have basic algebraic properties, an accurate description of QRM must necessarily specify the complete algebraic structure and symmetry transformation properties of the Hamiltonian. This means defining all the composite atom-light dynamical operators such as the composite excitation number, population inversion and related operators, which properly characterize the algebraic symmetry and dynamical properties generated by the QRM Hamiltonian. Specifying and using only one or an incomplete set of these composite dynamical operators just gives an algebraically limited approximate theory.

In particular, operators which generate symmetry transformations play a key role in determining the algebraic structure and internal dynamical properties of a system. Operators which commute with the Hamiltonian generate symmetry transformations which leave the system invariant. Such operators are useful in determining the exact eigenvalues and eigenstates of the system through diagonalization of the Hamiltonian. Equally important are operators generating symmetry transformations which determine the conjugates of the system. Besides providing additional insights into the structure of the system, such symmetry transformations often yield simpler forms of the system Hamiltonian which are exactly solvable or much easier to handle in fairly accurate approximations. The complete symmetry structure of a system is thus characterized by operators which generate invariance and conjugation transformations. It is not sufficient to define only operators which commute with the Hamiltonian and generate invariance transformations.

The underlying problem within the current theoretical framework of QRM may now be well explained. The great theoretical advances in studies of the symmetry properties of QRM have focussed attention only on symmetry transformation operators which commute with the Hamiltonian $H$ in equation (1$a$) or its fairly more general biased asymmetric form $H+\epsilon\sigma_x$ [3 , 4 , 17 , 20-23]. The parity symmetry operator
$\hat{\Pi}=e^{\pm i\pi(\hat{a}^\dagger\hat{a}+s_z)}$, appropriately expressed in the form $\hat{\Pi}=\sigma_z{\cal P}$
(${\cal P}=e^{-i\pi\hat{a}^\dagger\hat{a}}$), commutes with the basic QRM Hamiltonian $H$ [3 , 4 , 17 , 20], while the recently discovered generalized hidden symmetry operator ${\bf J}_\epsilon$ ($\epsilon={\frac{l}{2}} ~,~ l=0 , 1 , 2 , 3 , ...$) commutes with the asymmetric QRM Hamiltonian $H+\epsilon\sigma_x$ [21 , 22 , 23] where the biasing parameter $\epsilon$ takes integer and half-integer values as defined. It is established in [21 , 22 , 23] that the general form of the hidden symmetry operator includes the basic parity symmetry operator as the $l=0$ ($\epsilon=0)$ case, i.e., $\hat{\Pi}={\bf J}_0$. The parity and the generalized hidden symmetry operators generate invariance transformations and are useful in determining the exact eigenvalues and eigenstates of the QRM system through diagonalization of the Hamiltonian $H$ , $H+\epsilon\sigma_x$ [3 , 4 , 20-23]. In addition, the parity symmetry has been used in [6 , 17] and the related experiments [10 , 11] to gain insight into the internal dynamics of QRM, revealing fundamental quantum features in the ultra/deep strong coupling regimes.

Unfortunately, as explained earlier, the specification of only the parity and the generalized hidden symmetry operators does not define the complete symmetry structure and transformation properties of QRM. Other symmetry operators which generate transformations of the QRM Hamiltonian into its corresponding conjugates need also be specified. Indeed, it is established in this article that the parity operator is part of an algebraically closed  set of symmetry transformation operators $\hat{\Pi}_j=\sigma_j{\cal P}$, $j=z , y , x$, where $\hat{\Pi}_z$ is the parity symmetry operator, while $\hat{\Pi}_y$ , $\hat{\Pi}_x$ are identified as ``duality'' symmetry operators which generate symmetry conjugates of the QRM Hamiltonian. In particular, the duality symmetry transformations reveal an important algebraic property that corresponding JC and aJC operators are duality conjugates.

In progressing towards introduction of the complete set of parity and duality symmetry operators $\hat{\Pi}_j=\sigma_j{\cal P}$, $j=z , y , x$, it becomes important to note that the parity symmetry operator $\hat{\Pi}$ is generated by the JC excitation number operator $\hat{N}_{JC}=\hat{a}^\dagger\hat{a}+s_z$, generally considered to be the only excitation number operator in QRM used extensively in both theory and experiments [1 , 6 , 10 , 11 , 17], yet a simple symmetrization of the free evolution component $H_0$ of QRM Hamiltonian in equation (1$a$) in the form (~$2(aA+bB)=(a+b)(A+B)+(a-b)(A-B)$~)
$$H_0=\frac{1}{2}\hbar(\delta_+\hat{N}_{JC}+\delta_-\hat{N}_{aJC}) \ ; \quad\quad \hat{N}_{JC}=\hat{a}^\dagger\hat{a}+s_z \ ; \quad\quad \hat{N}_{aJC}=\hat{a}^\dagger\hat{a}-s_z \ ; \quad\quad \delta_\pm=\omega\pm\omega_0 \eqno(1b)$$
reveals that, besides the well known JC excitation number operator $\hat{N}_{JC}$, the algebraic structure of QRM is also characterized by another excitation number operator $\hat{N}_{aJC}$, identified as the anti-Jaynes-Cummings (aJC) excitation number operator, first constructed and proved conserved in aJC interaction in [23]. It is also proved explicitly in [23] that the aJC excitation number operator generates the same parity symmetry operator of QRM according to $\hat{\Pi}=e^{\pm i\pi\hat{N}_{JC}}=e^{\pm i\pi\hat{N}_{aJC}}$. In general, $\hat{N}_{JC}$ , $\hat{N}_{aJC}$ generate the respective $U(1)$ symmetry operators of the JC and aJC Hamiltonians. As demonstrated below, the two commuting excitation number operators are related by $\hat{\Pi}_y=\sigma_y{\cal P}$ , $\hat{\Pi}_x=\sigma_x{\cal P}$ symmetry transformations as duality conjugates and both must be specified in a complete algebraic structure of QRM.

For completeness, the interaction Hamiltonian $H_I$ in equation (1$a$) may also decomposed into a sum of JC and aJC components in the familiar form
$$H_I=H_{IJC}+H_{IaJC} \ ; \quad\quad H_{IJC}=\hbar g(\hat{a}^+s_-+\hat{a}s_+) \ ; \quad\quad H_{IaJC}=\hbar g(\hat{a}s_-+\hat{a}^+s_+) \eqno(1c)$$
so that substituting equations (1$b$) , (1$c$) into equation (1$a$) provides a decomposition of the QRM Hamiltonian into JC and aJC components in the form
$$H=H_{JC}+H_{aJC} \ ; \quad\quad H_{JC}=\frac{1}{2}\hbar(\delta_+\hat{N}_{JC}+H_{IJC} \ ; \quad\quad H_{aJC}=\frac{1}{2}\hbar(\delta_-\hat{N}_{aJC}+H_{IaJC} \eqno(1d)$$
In general, the JC and aJC components are found to be duality conjugates under the symmetry transformations generated by the operators $\hat{\Pi}_y$ , $\hat{\Pi}_x$.

It is now noted that, in addition to the excitation number operators which generate the parity symmetry operator, two other symmetry transformation operators which determine QRM ``duality'' symmetry conjugates exist as defined in this article. The ``duality'' symmetry transformations by both $\hat{\Pi}_y$ and $\hat{\Pi}-x$ map the JC ,  aJC excitation number operators into each other. For the JC , aJC interaction Hamiltonians, the operator $\hat{\Pi}_y$ generates a \emph{symmetric} duality transformation $H_{IJC}\to H_{IaJC}$ , $H_{IaJC}\to H_{JC}$, while the operator $\hat{\Pi}_x$ generates an \emph{antisymmetric} duality transformation $H_{IJC}\to-H_{IaJC}$ , $H_{IaJC}\to-H_{IJC}$. It then follows from equation (1$c$) that the duality symmetry transformations either map the full QRM interaction Hamiltonian $H_I$ onto itself ($H_I\to H_I$) or onto its mirror image ($H_I\to-H_I$), thus leading to characterization as symmetric or antisymmetric transformations. Based on the standard property that the sum or difference of an operator and its conjugate yields an effective operator, it is established that the sum (symmetric linear combination) of the QRM Hamiltonian $H$ and its symmetric duality conjugate yields an effective \emph{bosonic} Hamiltonian, while the difference (antisymmetric linear combination) with the antisymmetric duality conjugate yields an effective \emph{fermionic} Hamiltonian. A product of the duality symmetry operators provides a transform of the QRM Hamiltonian by changing only the sign of the interaction Hamiltonian $H_I$, so that the difference of the Hamiltonian and the transform provides a useful effective \emph{coupling-only} QRM Hamiltonian. It emerges that the effective bosonic, fermionic and coupling-only QRM Hamiltonians take familiar forms usually obtained as approximations in the general theoretical methods [16-19 , 25].

This article is organized as follows. QRM symmetry operators are introduced in section~$2$, where the transformation properties on the basic quantized light mode and atomic spin operators are presented. In section~$3$, the symmetry transformations are applied on the QRM Hamiltonian $H$ to determine invariance and duality conjugation properties. Basic dynamical properties of the symmetric and antisymmetric forms of QRM Hamiltonian are discussed briefly. Section~$4$ contains the Conclusion.

\section{Symmetry transformation operators}
A set of symmetry transformation operators $\hat{N}_j$, $j=z , y , x$ are introduced, defined in terms of the light mode excitation number $\hat{a}^\dagger\hat{a}$ and the atomic spin operators $s_j=\frac{1}{2}\sigma_j$ in the form
$$\hat{N}_z=\hat{a}^\dagger\hat{a}+s_z \ ; \quad\quad \hat{N}_y=\hat{a}^\dagger\hat{a}+s_y \ ; \quad\quad \hat{N}_x=\hat{a}^\dagger\hat{a}+s_x \eqno(2a)$$
where $\hat{N}_z$ is just the JC excitation number operator $\hat{N}_{JC}$ defined in the Introduction. These operators generate symmetry transformation operators of the general form
$$U_j(\theta)=e^{\pm i\theta\hat{N}_j} \ ; \quad j=z , y , x \eqno(2b)$$
where $\theta$ is taken real in these definitions. The special case $\theta=\pi$ provides the basic symmetry operators of interest in this article, which after substituting $\hat{N}_j$ from equation (2$a$) and reorganizing take the form
$$\hat{\Pi}_j=e^{\pm i\pi\hat{N}_j} \quad\quad\Rightarrow\quad\quad \hat{\Pi}_j=\sigma_j{\cal P} \ ; \quad\quad
{\cal P}=e^{\pm i\hat{a}^\dagger\hat{a}} \ ; \quad j=z , y , x \eqno(2c)$$
in the conventional representation adopted in [17 , 21 , 22 , 23], noting that for $j=z$, the operator $\hat{\Pi}_z$ is just the parity symmetry operator $\hat{\Pi}$ defined earlier. It is easily established that the operators $\hat{\Pi}_j$ as defined in equation (2$c$) satisfy a closed $SU(2)$ symmetry group algebra
$$\{\hat{\Pi}_j~,~\hat{\Pi}_k\}=2\delta_{jk} \ ; \quad\quad [~\hat{\Pi}_j~,~\hat{\Pi}_k]=2i\epsilon_{jkl}\hat{\Pi}_l{\cal P} \eqno(2d)$$
noting that ${\cal P}^2=I$ in the computational space. By introducing $\hat{\Pi}_\pm=\sigma_\pm{\cal P}$ ($\sigma_\pm=\sigma_x\pm i\sigma_y$), the $SU(2)$ Lie algebra of the closed set $\{~\hat{\Pi}_z~,~\hat{\Pi}_+~,~\hat{\Pi}_-\}$ easily provides the standard eigenvalues and eigenstates of the parity symmetry operator $\hat{\Pi}_z$.

The operators $\hat{\Pi}_j$ generate symmetry transformations of the light mode and atomic spin operators in the form
$$\hat{\Pi}_z^\dagger\hat{a}\hat{\Pi}_z=-\hat{a} \ ; \quad\quad \hat{\Pi}_z^\dagger\hat{a}^\dagger\hat{\Pi}_z=-\hat{a}^\dagger \ ; \quad\quad \hat{\Pi}_z^\dagger s_z\hat{\Pi}_z=s_z \ ; \quad\quad \hat{\Pi}_z^\dagger s_\mp\hat{\Pi}_z=-s_\mp \eqno(2e)$$
$$\hat{\Pi}_y^\dagger\hat{a}\hat{\Pi}_y=-\hat{a} \ ; \quad\quad \hat{\Pi}_y^\dagger\hat{a}^\dagger\hat{\Pi}_y=-\hat{a}^\dagger \ ; \quad\quad \hat{\Pi}_y^\dagger s_z\hat{\Pi}_y=-s_z \ ; \quad\quad \hat{\Pi}_y^\dagger s_\mp\hat{\Pi}_y=-s_\pm \eqno(2f)$$
$$\hat{\Pi}_x^\dagger\hat{a}\hat{\Pi}_x=-\hat{a} \ ; \quad\quad \hat{\Pi}_x^\dagger\hat{a}^\dagger\hat{\Pi}_x=-\hat{a}^\dagger \ ; \quad\quad \hat{\Pi}_x^\dagger s_z\hat{\Pi}_x=-s_z \ ; \quad\quad \hat{\Pi}_x^\dagger s_\mp\hat{\Pi}_x=s_\pm \eqno(2g)$$
The transformations in equation (2$e$) confirm that $\hat{\Pi}_z$ is the standard parity symmetry operator of QRM , which leaves the Hamiltonian $H$ in equation (1$a$) invariant. New QRM symmetry transformation properties are generated by the operators $\hat{\Pi}_y$ , $\hat{\Pi}_x$, which according to the actions on the atomic spin operators $s_\mp$ in equations (2$f$) , (2$g$), may be interpreted as ``duality'' symmetry operators.

\section{QRM parity and duality symmetry conjugates}
Applying the operator $\hat{\Pi}_z$ symmetry transformation, equivalent to the parity symmetry transformation, on the JC , aJC operators and the QRM Hamiltonian in equations (1$b$)-(1$d$), then using the relations obtained in equation (2$e$) leaves the operators invariant according to
$$\hat{\Pi}_z^\dagger~\hat{N}_{JC}~\hat{\Pi}_z=\hat{N}_{JC} \ ; \quad \hat{\Pi}_z^\dagger~\hat{N}_{aJC}~\hat{\Pi}_z=\hat{N}_{aJC} \ ; \quad\quad \hat{\Pi}_z^\dagger~H_{IJC}~\hat{\Pi}_z=H_{IJC} \ ; \quad \hat{\Pi}_z^\dagger~H_{IaJC}~\hat{\Pi}_z=H_{IaJC} $$
$$\hat{\Pi}_z^\dagger~H_0~\hat{\Pi}_z=H_0 \ ; \quad\quad \hat{\Pi}_z^\dagger~H_I~\hat{\Pi}_z=H_I \ ; \quad\quad \hat{\Pi}_z^\dagger~H~\hat{\Pi}_z=H \eqno(3a)$$
which provides the standard parity symmetry of QRM. The associated commutation relation $[~\hat{\Pi}_z~,~H~]=0$ follows easily from the invariance relation.

Applying the operators $\hat{\Pi}_y$ , $\hat{\Pi}_x$ transformations obtained in equations (2$f$) , (2$g$) on the excitation number operators and the free evolution Hamiltonian provides the duality symmetry transformations
$$j=y ~,~ x~:\quad\quad \hat{\Pi}_j^\dagger~\hat{N}_{JC}~\hat{\Pi}_j=\hat{N}_{aJC} \ ; \quad\quad \hat{\Pi}_j^\dagger~\hat{N}_{aJC}~\hat{\Pi}_j=\hat{N}_{JC} $$
$$\hat{\Pi}_j^\dagger~H_0~\hat{\Pi}_j=\overline{H}_0 \ ; \quad\quad \hat{\Pi}_j^\dagger~\overline{H}_0~\hat{\Pi}_j=H_0 \ ; \quad\quad \overline{H}_0=\frac{1}{2}\hbar(\delta_+\hat{N}_{aJC}+\delta_-\hat{N}_{JC}) \eqno(3b)$$
revealing that the JC and aJC excitation number operators $\hat{N}_{JC}$ , $\hat{N}_{aJC}$ are duality symmetry conjugates, while $\overline{H}_0$ is the duality symmetry conjugate of the free evolution Hamiltonian $H_0$.

Similarly, applying the operators $\hat{\Pi}_y$ , $\hat{\Pi}_x$ transformations on the JC , aJC interaction Hamiltonians provides the duality symmetry transformations
$$\hat{\Pi}_y^\dagger~H_{IJC}~\hat{\Pi}_y=H_{IaJC} \ ; \quad\quad \hat{\Pi}_y^\dagger~H_{IaJC}~\hat{\Pi}_y=H_{IJC} $$
$$\hat{\Pi}_x^\dagger~H_{IJC}~\hat{\Pi}_x=-H_{IaJC} \ ; \quad\quad \hat{\Pi}_x^\dagger~H_{IaJC}~\hat{\Pi}_x=-H_{IJC} \eqno(3c)$$
The JC and aJC interaction Hamiltonians, which have generally been interpreted as the rotating and counter-rotating components of the QRM Hamiltonian, are here identified as duality symmetry conjugates. It is then important to note that, just as the free evolution Hamiltonian $H_0$ is a sum of duality symmetry conjugates according to equations (1$b$) , (3$b$), the interaction Hamiltonian $H_I$ is also a sum of duality symmetry conjugates according to equations (1$c$) , (3$c$).

It emerges here that both duality symmetry operators $\hat{\Pi}_y$ , $\hat{\Pi}_x$ transform $H_0$ , $\overline{H}_0$ directly into each other ($H_0\leftrightarrow\overline{H}_0$) without sign differences ($\pm$). However, it follows from the relations in equations (1$c$) , (3$c$) that $\hat{\Pi}_y$ maps the interaction Hamiltonian $H_I$ onto itself ($H_I\to H_I$), while $\hat{\Pi}_x$ generates the mirror image transformation $H_I\to-H_I$. This property leads to an interpretation that $\hat{\Pi}_y$ generates symmetric and $\hat{\Pi}_x$  antisymmetric duality symmetry conjugations of QRM, which are treated separately.

\subsection{Symmetric QRM duality conjugation}
Applying the $\hat{\Pi}_y$ transformation on $H_I$ in equation (1$c$) and using the corresponding relations from equation (3$c$) generates the symmetric duality transformation
$$\hat{\Pi}_y^\dagger~H_I~\hat{\Pi}_y=H_I \quad\quad\Rightarrow\quad\quad [~\hat{\Pi}_y~,~H_I]=0 \eqno(4a)$$
In a standard interpretation, the symmetric duality transformation, governed by the specified commutation relation, leaves the interaction Hamiltonian $H_I$ invariant. It follows from equations (3$b$) , (4$a$) that application of $\hat{\Pi}_y$ transforms the QRM Hamiltonian $H$ in equation (1$a$) into its symmetric duality conjugate $\overline{H}_+$ according to
$$\hat{\Pi}_y^\dagger~H~\hat{\Pi}_y=\overline{H}_+ \ ; \quad\quad \hat{\Pi}_y^\dagger~\overline{H}_+~\hat{\Pi}_y=H \ ; \quad\quad \overline{H}_+=\overline{H}_0+H_I \eqno(4b)$$

\subsubsection{Symmetric QRM Hamiltonian : effective bosonic dynamics}
Taking a symmetric linear combination of the QRM Hamiltonian $H$ in equation (1$a$) and its symmetric duality conjugate $\overline{H}_+$ in equation (4$b$), then using the definitions from equations (1$a$) , (1$b$) , (3$b$) , (4$b$), provides the symmetric QRM Hamiltonian $H_+$ obtained as
$$H_+=\frac{1}{2}(H+\overline{H}_+) \quad\quad\Rightarrow\quad\quad H_+=\hbar\omega\hat{a}^\dagger\hat{a}+\hbar g(\hat{a}+\hat{a}^\dagger)\sigma_x \eqno(5a)$$
after introducing the Pauli spin operator $\sigma_x=s_-+s_+$ to give a familiar form. This form shows that the symmetric QRM Hamiltonian $H_+$ is an effective bosonic Hamiltonian describing the dynamics of the quantized light mode driven by the atomic spin-dependent force. Notice that $H_+$ is invariant under the symmetric $\hat{\Pi}_y$ duality symmetry transformation according to
$$\hat{\Pi}_y^\dagger~H_+\hat{\Pi}_y=H_+ \quad\quad\Rightarrow\quad\quad [~\hat{\Pi}_y~,~H_+~]=0 \eqno(5b)$$
The effective bosonic Hamiltonian $H_+$ first arose as an exactly solvable degenerate spin state ($\omega_0=0$) approximation of QRM in [17] and has inspired in-depth theoretical and experimental investigations of QRM dynamics under spin-dependent forces over the entire coupling parameter range [1 , 2 , 4 , 10 , 11 , 18 , 19 , 25-31]. The simple derivation through symmetric duality transformations in equations (4$b$) , (5$a$) reveals the important physical property that the effective bosonic Hamiltonian $H_+$ is an exact symmetric QRM Hamiltonian, entirely independent of the atomic spin state transition angular frequency $\omega_0$, thus applicable to both degenerate ($\omega_0=0$) and non-degenerate ($\omega_0\ne0$) spin state cases.

The bosonic nature of $H_+$ is easily demonstrated by introducing composite hermitian conjugate operators $\hat{b}$ , $\hat{b}^\dagger$ satisfying bosonic algebra according to
$$\hat{b}=\hat{a}\sigma_x \ ; \quad \hat{b}^\dagger=\hat{a}^\dagger\sigma_x \ ; \quad \sigma_x^2=I~:\quad\quad [~\hat{b}~,~\hat{b}^\dagger~]=1 \ ; \quad\quad [~\hat{b}~,~\hat{b}~]=0 \ ; \quad\quad [~\hat{b}^\dagger~,~\hat{b}^\dagger~]=0 \eqno(5c)$$
where $I$ is the $2\times2$ identity matrix. The Hamiltonian in equation (5$a$) now takes the form
$$H_+=\hbar\omega\hat{b}^\dagger\hat{b}+\hbar g(\hat{b}+\hat{b}^\dagger) \eqno(5d)$$
Application of the Heisenberg equation of motion for the operator $\hat{b}$ gives time evolution reorganized in the appropriate form
$$\hat{b}(t)=e^{-i\omega t}\left(\hat{b}+\frac{g}{\omega}(1-e^{i\omega t})\right) \ ; \quad\quad \hat{b}=\hat{b}(0) \eqno(5e)$$
which is a displaced bosonic state annihilation operator easily obtained through a time evolution operator $U(t)$ according to
$$U(t)=e^{-i\omega t\hat{b}^\dagger\hat{b}}D(\beta(t)) \ ; \quad D(\beta(t))=e^{\beta(t)\hat{b}^\dagger-\beta^*(t)\hat{b}} \ ; \quad \beta(t)=\frac{g}{\omega}(1-e^{i\omega t}) \ ; \quad\quad \hat{b}(t)=U^\dagger(t)~\hat{b}~U(t) \eqno(5f)$$
noting
$$U^\dagger(t)=D^\dagger(\beta(t))e^{i\omega t\hat{b}^\dagger\hat{b}}~: \quad\quad
D^\dagger(\beta(t))e^{i\omega t\hat{b}^\dagger\hat{b}}\hat{b}e^{-i\omega t\hat{b}^\dagger\hat{b}}D(\beta(t))=
e^{-i\omega t}D^\dagger(\beta(t))\hat{b}D(\beta(t)) \eqno(5g)$$
Determination of the time evolution operator $U(t)$ in equation (5$f$) provides the exact solution of the dynamics generated by the bosonic QRM Hamiltonian $H_+$. Taking the atom initially in the excited state $\vert e\rangle$, the light mode in the vacuum state $\vert 0\rangle$ and introducing the $\sigma_x$ eigenstates $\vert\pm\rangle=\frac{1}{\sqrt{2}}(\vert e\rangle\pm\vert g\rangle)$ with $\sigma_x\vert\pm\rangle=\pm\vert\pm\rangle$, the general time evolving state $\vert\Psi(t)\rangle=U(t)\vert e0\rangle$ is obtained as an entangled atom-light state in the form
$$\vert\Psi(t)\rangle=\vert\beta_+(t)\rangle\vert e\rangle+\vert\beta_-(t)\rangle\vert g\rangle \ ; \quad\quad  \vert\beta_\pm(t)\rangle=\frac{1}{\sqrt{2}}(\vert\beta(t)\rangle\pm\vert -\beta(t)\rangle) \eqno(5h)$$
where $\vert\beta_\pm(t)\rangle$ as defined is the coherent light mode Schroedinger cat state. The time evolving state $\vert\Psi(t)\rangle$ of the effective bosonic Hamiltonian is therefore an entanglement of the coherent light Schroedinger cat and atomic spin states. The physical interpretation is that in the effective bosonic dynamics driven by the atomic spin-dependent force, the dynamical evolution of the full quantum Rabi model is described by the general time evolving entangled coherent light Schroedinger cat and atomic spin state $\vert\Psi(t)\rangle$ in equation (5$h$). The appropriate order parameter for studying the dynamical evolution is the mean value of the bosonic excitation number operator $\hat{b}^\dagger\hat{b}=\hat{a}^\dagger\hat{a}$ in the state $\vert\Psi(t)\rangle$. The fundamental quantum properties and practical applications of QRM dynamics described by these nonclassical bosonic atom-light entangled states are well established in the theoretical and experimental studies [1 , 4 , 6 , 10 , 17 , 19 , 26] where details which need not be repeated here can be found.

\subsection{Antisymmetric QRM duality conjugation}
Applying the $\hat{\Pi}_x$ transformation on $H_I$ in equation (1$c$) and using the corresponding relations from equation (3$c$) generates the antisymmetric duality transformation
$$\hat{\Pi}_x^\dagger~H_I~\hat{\Pi}_x=-H_I \quad\quad\Rightarrow\quad\quad \{~\hat{\Pi}_x~,~H_I\}=0 \eqno(6a)$$
In an interpretation, the antisymmetric duality transformation, governed by the specified anticommutation relation, leaves the operator form of the interaction Hamiltonian $H_I$ invariant, but maps it onto its mirror image. It follows from equations (3$b$) , (6$a$) that application of $\hat{\Pi}_x$ transformation on the QRM Hamiltonian $H$ in equation (1$a$) generates the antisymmetric duality conjugate $\overline{H}_-$ according to
$$\hat{\Pi}_x^\dagger~H~\hat{\Pi}_x=\overline{H}_- \ ; \quad\quad \hat{\Pi}_x^\dagger~\overline{H}_-~\hat{\Pi}_x=H \ ; \quad\quad \overline{H}_-=\overline{H}_0-H_I \eqno(6b)$$

\subsubsection{Antisymmetric QRM Hamiltonian : effective fermionic dynamics}
Taking an antisymmetric linear combination of the QRM Hamiltonian $H$ in equation (1$a$) and its antisymmetric duality conjugate $\overline{H}_-$ in equation (6$b$), then using the definitions from equations (1$a$) , (1$b$) , (3$b$) , (6$b$), provides the antisymmetric QRM Hamiltonian $H_-$ obtained as
$$H_-=\frac{1}{2}(H-\overline{H}_-) \quad\quad\Rightarrow\quad\quad H_-=\hbar\omega_0s_z+\hbar g(\hat{a}+\hat{a}^\dagger)\sigma_x \eqno(7a)$$
This form shows that the antisymmetric QRM Hamiltonian $H_-$ is an effective fermionic Hamiltonian describing the dynamics of the atomic spin driven by the quantized light mode quadrature-dependent force. The QRM fermionic Hamiltonian $H_-$ maps onto its mirror image under the antisymmetric $\hat{\Pi}_x$ duality transformation according to
$$\hat{\Pi}_x^\dagger~H_-~\hat{\Pi}_x=-H_- \quad\quad\Rightarrow\quad\quad \{~\hat{\Pi}_x~,~H_-~\}=0 \eqno(7b)$$
It is remarkable that the QRM fermionic Hamiltonian $H_-$ in equation (7$a$) is similar to the one-dimensional Dirac Hamiltonian for a fermion in relativistic quantum mechanics, which has inspired quantum simulations of the Dirac equation with trapped ions in quantum optics [1 , 31 , 32].

Unlike the corresponding bosonic Hamiltonian $H_+$ in equation (5$a$) which is expressible in terms of composite operators satisfying bosonic algebra according to equations (5$c$) , (5$d$), the QRM fermionic Hamiltonian $H_-$ with a free evolution spin-only component in equation (7$a$) is not expressible in terms of composite atom-light operators satisfying fermionic algebra characterizing the antisymmetric duality transformation in equations (6$a$) , (7$b$). Consequently, exact analytical solutions of the dynamical evolution generated by $H_-$ have proved too difficult to determine and only approximate solutions, exemplified by the detailed analysis in [25], have been provided in the quantum optics literature.

Corresponding to the bosonic case $H_+$ characterized by $\omega\ne0$ , $\omega_0=0$ in [25] where the driving spin operator $\sigma_x$ is averaged in its eigenstate $\vert\pm\rangle$ and replaced by an eigenvalue, the authors applied a similar procedure in the analysis of the dynamical features of the ferminic Hamiltonian $H_-$ characterized there by $\omega=0$ , $\omega_0\ne0$ by replacing the driving light mode quadrature operator $\hat{x}=\hat{a}+\hat{a}^\dagger$ with a mean value $x$ in an eigenstate of the light mode. The resulting time evolution equation is then exactly solvable if $x$ is time-independent, but remains challenging if $x$ is time-dependent.

In the present article, the property that the equations of dynamics generated by the correlated bosonic and fermionic forms $H_+$ , $H_-$ (~$[~H_+~,~H_-~]\ne0$~) of QRM are solved simultaneously means that the driving light mode quadrature operator $\hat{a}+\hat{a}^\dagger$ in $H_-$ may be replaced with its mean value evaluated in the entangled state $\vert\Psi(t)\rangle$ in the bosonic dynamics generated by $H_+$ in equation (5$h$), so that, as in [25], the fermionic Hamiltonian takes the effective form
$${\cal H}_-=\hbar\omega_0s_z+2\hbar gx(t)\sigma_x \ ; \quad\quad x(t)=\langle\Psi(t)\vert\hat{a}+\hat{a}^\dagger\vert\Psi(t)\rangle \eqno(7c)$$
The general time evolving state vector describing the dynamics of the fermionic system may be obtained through diagonalization of the effective Hamiltonian ${\cal H}_-$, carefully taking account of the time-dependence of the light mode mean quadrature $x(t)$. Details of such analysis can be found in [25].

Another possible method of solution is based on the important property that the fermionic Hamiltonian $H_-$ in equation (7$a$) is invariant under the parity symmetry transformation. The elaborate Braak methods developed in [3 , 4 , 20 , 37] can be applied to obtain exact solutions of the eigenvalue equation for $H_-$ in the Bargmann space, which is simpler, but has not been considered at all, essentially due to the integrability of the full QRM Hamiltonian $H$. The resulting eigenvalue and eigenstate spectrum can be used to determine dynamical evolution generated by $H_-$ using methods developed in [5 , 6].

In general, the time evolving state of the effective fermionic Hamiltonian $H_-$ is a generalized atomic spin coherent state. The physical interpretation then follows that, in the effective fermionic dynamics driven by the light mode quadrature-dependent force, the dynamical evolution of the full quantum Rabi model is described by time evolving generalized atomic spin coherent states.

\subsection{$\hat{\Pi}_{yx}$ symmetry transformation : effective coupling-only dynamics}
Another interesting symmetry transformation operator is obtained as a product of the operators $\hat{\Pi}_y$ , $\hat{\Pi}_x$ in either order with $\hat{\Pi}_y$ to the left or right of $\hat{\Pi}_x$ denoted by $\hat{\Pi}_{yx}$ or $\hat{\Pi}_{xy}$ according to
$$\hat{\Pi}_{yx}=\hat{\Pi}_y\hat{\Pi}_x \ ; \quad\quad \hat{\Pi}_{xy}=\hat{\Pi}_x\hat{\Pi}_y \eqno(8a)$$
Applying $\hat{\Pi}_{yx}$ or $\hat{\Pi}_{xy}$ symmetry transformation on the QRM Hamiltonian $H$ and using the relations obtained in equations (3$b$) , (4$a$) , (4$b$) , (6$a$) , (6$b$) provides the transform $\widetilde{H}$ of the QRM Hamiltonian in the form
$$\hat{\Pi}_{yx}^\dagger~H~\hat{\Pi}_{yx}=\widetilde{H} \ ; \quad\quad \hat{\Pi}_{yx}^\dagger~\widetilde{H}~\hat{\Pi}_{yx}=H \ ; \quad\quad \widetilde{H}=H_0-H_I \eqno(8b)$$
noting that the $\hat{\Pi}_{xy}$ symmetry transformation gives the same result. Substituting $H_0$ , $H_I$ as defined in equation (1$a$) reveals that the QRM Hamiltonian transform $\widetilde{H}$ has been used as an alternative form for describing QRM dynamics in [34] where the quantum phase transition phenomenon is presented. It follows from equation (8$b$) that the two forms of QRM Hamiltonian are related by symmetry transformation.

Taking an antisymmetric linear combination of $H$ in equation (1$a$) and its transform $\widetilde{H}$ in equation (8$b$) provides an effective coupling-only Hamiltonian $H_{RI}$ in the form
$$H_{RI}=\frac{1}{2}(H-\widetilde{H}) \quad\quad\Rightarrow\quad\quad H_{RI}=H_I \eqno(8c)$$
where the QRM interaction Hamiltonian $H_I$ is defined in equation (1$a$). The exactly solvable dynamics of the coupling-only QRM Hamiltonian $H_{RI}$ has been widely used to generate Schroedinger cat states [27 , 35 , 36], where it is characterized as $\omega=0$ , $\omega_0=0$ approximation of QRM Hamiltonian in the ultra/deep strong coupling regime. The time evolution operator for dynamics generated by $H_{RI}$ follows as a simple solution of the time-dependent Schroedinger equation in the form
$$U_{RI}(t)=e^{\overline{\beta}(t)(\hat{b}+\hat{b}^\dagger)} \ ; \quad\quad \overline{\beta}(t)=-igt \eqno(8d)$$
where the conjugate composite atom-light bosonic operators $\hat{b}$ , $\hat{b}^\dagger$ are defined in equation (5$c$). The time evolving state of the effective coupling-only Hamiltonian is thus an entanglement of the coherent light Schroedinger cat and atomic spin states. The time evolving states of the effective bosonic and coupling-only Hamiltonians differ only in the form of the respective displacement parameters $\beta(t)$ in equation (5$f$) and $\overline{\beta}(t)$ in equation (8$d$). The physical interpretation follows that, in the effective coupling-only dynamics generated only by the coupled light mode quadrature and atomic spin forces, the dynamical evolution of the full quantum Rabi model is described by a general time evolving entangled coherent light Schroedinger cat and atomic spin state. General dynamical features described by the Schroedinger cat states generated by $U_{RI}$ with displacement variable $\overline{\beta}(t)$ exactly in the form defined in equation (8$d$) are discussed in [35], agreeing also with the results in [27 , 36] for phase $\phi=0$. Details which need not be repeated here can be found in the cited articles.

\section{Conclusion}
A complete set of symmetry transformation operators comprising parity and duality symmetry operators have been defined within the general algebraic structure of QRM. The operators satisfy a closed $SU(2)$ symmetry group algebra. The duality symmetry operators transform the JC and aJC excitation number operators into each other as duality symmetry conjugates, while the corresponding JC and aJC interaction Hamiltonians transform into each other as symmetric or antisymmetric duality symmetry conjugates. The parity symmetry transformation leaves the QRM Hamiltonian invariant, while the duality symmetry transformations generate symmetric and antisymmetric conjugates of the Hamiltonian. The transformation generated by the product of the two duality symmetry operators provides a QRM Hamiltonian transform differing only in the sign of the interaction component. The closed $SU(2)$ Lie algebra of the symmetry operators is useful in determining the standard eigenvalues and eigenstates of the parity symmetry operator. Symmetric or antisymmetric linear combinations of the QRM Hamiltonian and the respective symmetric or antisymmetric conjugates produce exact effective bosonic , fermionic or coupling-only Hamiltonians, which provide a clear physical picture of the internal dynamics of QRM. The emerging physical interpretation is that, the effective bosonic or coupling-only Hamiltonian describes the dynamics of the quantized light mode driven by a spin-dependent force, while the fermionic Hamiltonian describes the dynamics of the atomic spin driven by a quantized light mode quadrature-dependent force. The effective bosonic or coupling-only Hamiltonian generates a time evolving entangled coherent light Schroedinger cat and atomic spin state, while the effective fermionic Hamiltonian generates a time evolving generalized atomic spin coherent state. The important physical interpretation which arises is that the general dynamical evolution of the full quantum Rabi model is described by the nonclassical time evolving entangled coherent light Schroedinger cat and atomic states or generalized atomic spin coherent states. These nonclassical states have been prepared and used in developing quantum technology applications in a number of experiments. The remarkable achievement of the present work is that the effective bosonic, fermionic or coupling-only QRM Hamiltonians generated through the symmetry transformations are exact, not involving approximations based on the atomic spin and light mode angular frequencies or coupling strength as generally assumed in the earlier theoretical and experimental studies of QRM. Consequently, the brief exact results provided in this article may be considered to apply over the entire frequency and coupling parameter ranges. It is interesting that within the framework of the duality symmetry transformations, the full QRM dynamics has been developed without decomposition into coupling-strength related Jaynes-Cummings and anti-Jaynes-Cummings interaction mechanisms, each of which usually requiring serious approximations.

\section{Acknowledgement}
I thank Maseno University for providing facilities and a conducive work environment during the preparation of the manuscript. My colleague Chris Mayero has provided useful technical support.

\end{document}